\documentclass[twocolumn,prl,aps,superscriptaddress,showpacs,amsmath,amssymb,floatfix]{revtex4-1}
\usepackage{graphicx}
\usepackage{amssymb}
\usepackage{amsmath}
\usepackage{epsfig}
\usepackage{color}
\usepackage{mathtools}
\usepackage[colorlinks,linkcolor=blue,anchorcolor=blue,citecolor=blue,urlcolor=blue]{hyperref}
\usepackage{physics}

\begin{document}

\title{Gapped topological Fulde-Ferrell-Larkin-Ovchinnikov superfluids with artificial gauge potential and weak interaction}
\author{Yue-Xin Huang}
\affiliation{Key Lab of Quantum Information, Chinese Academy of Sciences, School of physics, University of Science and Technology of China, Hefei, 230026, P.R. China}
\author{Wei-Feng Zhuang}
\affiliation{Key Lab of Quantum Information, Chinese Academy of Sciences, School of physics, University of Science and Technology of China, Hefei, 230026, P.R. China}
\author{Zhen Zheng}
\affiliation{Key Lab of Quantum Information, Chinese Academy of Sciences, School of physics, University of Science and Technology of China, Hefei, 230026, P.R. China}
\author{Guang-Can Guo}
\affiliation{Key Lab of Quantum Information, Chinese Academy of Sciences, School of physics, University of Science and Technology of China, Hefei, 230026, P.R. China}
\affiliation{Synergetic Innovation Center of Quantum Information and Quantum Physics, University of Science and Technology of China, Hefei, 230026, P.R. China}
\author{Ming Gong}
\email{gongm@ustc.edu.cn}
\affiliation{Key Lab of Quantum Information, Chinese Academy of Sciences, School of physics, University of Science and Technology of China, Hefei, 230026, P.R. China}
\affiliation{Synergetic Innovation Center of Quantum Information and Quantum Physics, University of Science and Technology of China, Hefei, 230026, P.R. China}
\date{\today }

\begin{abstract}
The topological superfluids with Majorana zero modes have not yet been realized in ultracold atoms with Rashba spin-orbit coupling. Here 
we show that these phases can be realized with an artificial gauge potential, which can be regarded as a site-dependent rotating Zeeman
field. This potential breaks the inversion symmetry and plays the same role as Rashba spin-orbit coupling. In the inverted bands, this model can open a proper 
parameter regime for topological superfluids. Strikingly, we find that the interaction near the Fermi surface is dominated by the dispersion scattering in the same band, 
thus can realize topological phase with much weaker attractive interaction, as compared with the model with Rashba spin-orbit coupling. We find a large regime 
for the gapped topological Fulde-Ferrell-Larkin-Ovchinnikov superfluids and unveil the phase diagram with mean-field theory, which should be credible in the weak 
interaction regime. In regarding the negligible heating effect in realizing this potential in alkaline and rare-earth atoms, our model has the potential to be 
the first system to realize the long-sought topological FFLO phase and the associated Majorana zero modes.
\end{abstract}
\maketitle

The spin-orbit coupling (SOC) plays an important role in many physics in condensed matters \cite{hasan_colloquium_2010,qi_topological_2011, sinova2015spin, xiao2010berry}
and ultracold atoms \cite{galitski_spin-orbit_2013,stuhl_visualizing_2015,zhai_degenerate_2015,fu_production_2013,wall_synthetic_2016,kolkowitz_spinorbit-coupled_2017},
especially for the realization of topological phases and associated protected edge modes \cite{kitaev_unpaired_2001,tewari_quantum_2007,sato_non-abelian_2009,sarma_majorana_2015}. 
In solid materials, the SOC can naturally exist for systems without inversion symmetry \cite{sinova2015spin, xiao2010berry}. In ultracold atoms, this term can be induced using Raman couplings \cite{zhu_spin_2006,spielman_raman_2009,liu_effect_2009,anderson_synthetic_2012,zhai_degenerate_2015}. 
In recent years, both the one dimensional SOC \cite{lin_spin-orbit-coupled_2011,
zhang_collective_2012, wang_spin-orbit_2012,hamner_dicke-type_2014,ji_experimental_2014,olson_tunable_2014,li_spin-orbit_2016} and 
the two dimensional SOC have been realized in both fermionic \cite{huang_experimental_2016} and bosonic atoms \cite{wu_realization_2016}. 
These progresses have opened a new avenue in experiments to explore various exotic topological phases \cite{hasan_colloquium_2010, sinova2015spin, xiao2010berry, qi_topological_2011,fu_superconducting_2008,
cooper_stable_2009,oreg_helical_2010,lutchyn_majorana_2010,zhou_topological_2011,mourik_signatures_2012}.

In these experiments, the properties of the superfluids depends strongly on the direction of Zeeman field. For out-of-plane Zeeman field, which
opens an gap at zero momentum but preserves the inversion symmetry, Bardeen-Cooper-Schrieffer pairing is preferred \cite{cooper_bound_1956,bardeen_microscopic_1957}; 
however for the in-plane Zeeman field without inversion symmetry, the pairs may carry finite momentum for Fulde-Ferrell-Larkin-Ovchinnikov (FFLO) phase
\cite{fulde_superconductivity_1964,larkin_inhomogeneous_1964,loh_detecting_2010,liao_spin-imbalance_2010,zhang_topological_2013, wu_unconventional_2013,qu_fulde-ferrell-larkin-ovchinnikov_2014}. 
The fate of FFLO phase is still elusive in condensed matter physics and was one of the major concerns in ultracold atoms \cite{liao_spin-imbalance_2010}. The problem is that this phase
is very easy to enter the gapless regime due to band tiling effect \cite{chan_pairing_2014, liu_majorana_2014, zheng_fflo_2014}, thus not all FFLO phases can support topological protected 
Majorana zero modes. 

\begin{figure}[htpb]
    \centering
    \includegraphics[width=0.49\textwidth]{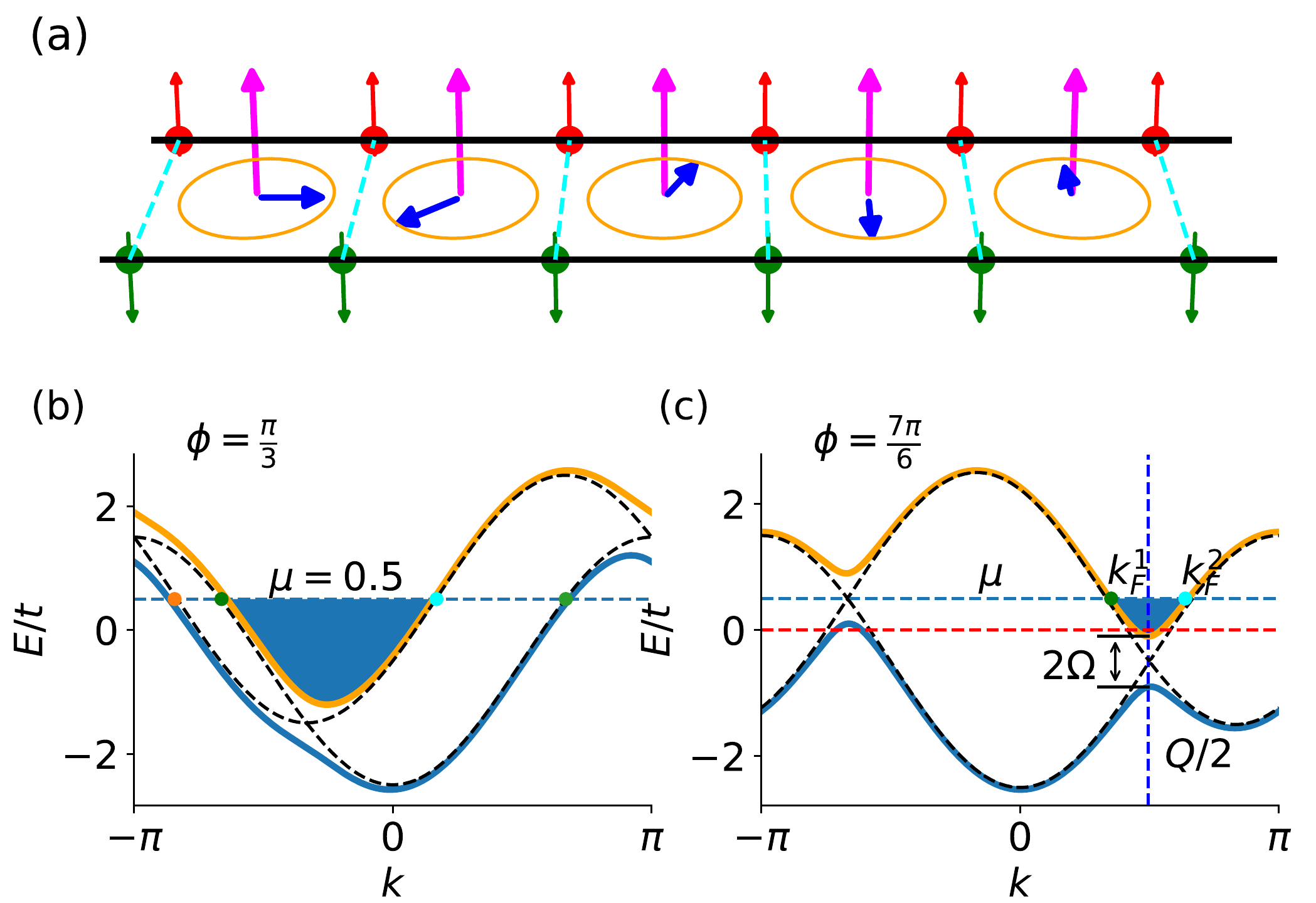}
    \caption{(Color online) (a) Topological superfluids in a one dimensional tight-binding model with artificial gauge potential. In this ladder representation the 
	two chain represent two different spins with the same spatial wave function. The vertical arrows and horizontal arrows represent the in-plane 
    and out-of-plane Zeeman field, respectively. The in-plane Zeeman field is induced by a site-dependent rotating Zeeman field. (b) and (c) Typical single 
    particle band structure for $\phi < \pi$ and $\phi > \pi$ (case for inverted band), respectively. The dashed and solid lines represent the cases 
    without and with $h$. Other parameters are $h=0.5t$, $\Omega=0.4t$.}
    \label{fig-fig1}
\end{figure}

In this work, we propose to realize the fully gapped topological FFLO phase and associated Majorana zero modes \cite{chen_inhomogeneous_2013,qu_fulde-ferrell-larkin-ovchinnikov_2014,mayaffre_evidence_2014,mizushima_topological_2005} based on only artificial gauge potential. This system has a number of interesting features
that are totally different from the previous proposals with Rashba SOC \cite{lin_spin-orbit-coupled_2011,
zhang_collective_2012, wang_spin-orbit_2012,hamner_dicke-type_2014,ji_experimental_2014,olson_tunable_2014,li_spin-orbit_2016}. In our model, the spin-momentum locking effect is 
realized by a site-dependent rotating Zeeman field, which can play the same role as Rashba SOC in momentum space.  
In the case of inverted band, the dispersion scattering near the Fermi surface is dominated among all possible scattering channels, thus it has the 
advantage of realizing topological FFLO superfluids with much weaker attractive interaction, as compared to that with Rashba SOC. 
In this regime, the mean-field theory should be credible, with which we map out the whole phase diagram. This model is relevant to the recent realized artificial gauge potentials 
in alkaline and rare-earth atoms with negligible heat effect \cite{dalibard_colloquium_2011,jaksch_creation_2003,osterloh_cold_2005, lin_synthetic_2009,celi_synthetic_2014,
goldman_light-induced_2014,mancini2015observation}.

We consider the following one dimensional model in an optical lattice (see the ladder representation in Fig. \ref{fig-fig1}a),
\begin{eqnarray}
        H=\mathcal{H}_0 + \mathcal{H}_\text{g} + V_\text{int}. 
            \label{eq-H}
\end{eqnarray}
In the first term, we consider the spin-independent hopping between the neighboring sites,
\begin{eqnarray}
        \mathcal{H}_0 &=&  - \sum_{m, s} (ta_{m,s}^\dagger a_{m+1,s} + t a_{m+1,s}^\dagger a_{m,s}+\mu n_{m,s})
    \nonumber\\
                &&+ h\sum_m (  n_{m\uparrow} -n_{m\downarrow}),
\end{eqnarray}
where $t$, $\mu$ and $h$ represent the tunneling, chemical potential and out-of-plane Zeeman field, respectively, $a_{m,s}^\dagger$
($a_{m,s}$) is the creation (annihilation) operator for Fermion particles at lattice site $m$ with spin $s = \uparrow,\downarrow$, and 
$n_{m,s} = a_{m,s}^\dagger a_{m,s}$. The artificial gauge potential reads as
\begin{equation}
        \mathcal{H}_\text{g} = \sum_m \left( \Omega e^{-im\phi} a_{m,\uparrow}^\dagger a_{m,\downarrow}
                +\Omega e^{im\phi} a_{m,\downarrow}^\dagger a_{m,\uparrow} \right),
                    \label{eq-Hg}
\end{equation}
where $\phi$ can be viewed as flux per plaquette in Fig. \ref{fig-fig1}a. 
In an optical lattice, this potential can be realized using different schemes, including laser assisted tunneling \cite{aidelsburger2013realization,miyake2013realizing,kennedy2015observation}
and driven optical lattice \cite{struckengineering, mancini2015observation, tai2017microscopy}. Very recently, this interaction has also been realized in rare-earth atoms
\cite{daley_quantum_2008,wall_synthetic_2016,kolkowitz_spinorbit-coupled_2017, livi2016synthetic}. This potential can lead to various applications
\cite{moller_composite_2009,ruostekoski_optical_2009,cooper_measuring_2010,dubcek_weyl_2015,zheng_artificial_2017}.
In these schemes, due to lacking of spontaneous emission in the large detuning limit, the heating effect is negligible. We notice that this potential can be regarded as a 
site-dependent rotating Zeeman field, which breaks the inversion symmetry and induces spin flipping, thus
can play the same role as Rashba SOC, although their mechanisms and realizations are totally different. In following we will show that these two methods can be regarded as 
``complementary'' mechanism to each other.

The phase carried by the Zeeman field can be gauged out by a transformation $a_{m,\uparrow}\rightarrow a_{m,\uparrow} e^{-i m \phi}$, while the
spin-down component is unchanged. Then the above single particle Hamiltonian can be written as
\begin{equation}
        \mathcal{H}(k) = \mathcal{H}_0 + \mathcal{H}_\text{g} = \epsilon_k + {\bf h}(k) \cdot \boldsymbol{\sigma}, 
\end{equation}
where ${\bf h} = (h_x, h_y, h_z)$, with $\boldsymbol{\sigma} = (\sigma_x, \sigma_y, \sigma_z)$ being the Pauli matrices. In above equation, we have
\begin{eqnarray} 
        \epsilon_k && = -t (\cos(k + \phi) + \cos(k)) - \mu, \quad \quad h_x = \Omega, \nonumber \\
             h_y && = 0, \quad h_z = h -t (\cos(k + \phi) - \cos(k)).
\end{eqnarray}
Now the flux phase carried by each plaquette enters the diagonal term, which induces misalignment between the two bands. Two typical band structures 
($\phi > \pi$ for inverted band) are presented in Fig. \ref{fig-fig1}b-c. In this model, the momentum dependent terms all appear in the diagonal term and the 
effective Zeeman field $h_z$, now, is momentum dependent, which, in the presence of in-plane Zeeman fields $h_x$ and $h_y$, will induce spin-momentum locking effect. 
This is different from the Rashba SOC investigated in literatures \cite{sinova2015spin, xiao2010berry}, in which not $h_z$, but $h_x$ and $h_y$ are odd function of $k$.
These two models can not be connected by basis rotation, so this new SOC term can be regarded as a ``complementary'' mechanism of Rashba SOC. 
For this reason, the artificial gauge potential can play the same role as the 
Rashba SOC. Notice that in ultracold atoms, the Rashba coupling strength is determined by the momentum of light, and can not be tuned easily in experiments. By fast modulating 
the Zeeman field, the effective SOC strength can only be tuned to a weaker value \cite{jimenez-garcia_tunable_2015,meng_experimental_2016, hamner2014dicke}. In our model, 
the ``effective'' SOC is induced by the cooperation of phase and Zeeman fields, thus can be tuned much easier in experiments.

We next consider the effect of weak attractive interactions on the properties of the superfluids, in which regime the mean field prediction is credible. 
For the inverted band presented in Fig. \ref{fig-fig1}c, it is possible to choose a proper $\mu$, which occupy the whole 
lower band and partially the higher band. The tilted band structure means that the mean momentum of the two Fermi points is nonzero. 
Then the weak attractive interaction between atoms should induce pairings near the two Fermi points, which obviously carry a finite momentum 
$Q \simeq (k_{\rm F}^1+ k_{\rm F}^2)$. Notice that the spin-momentum locking effect ensures that in each band the
eigenvectors contains both the spin up and down components, thus the pairing at the same band, which is strictly forbidden due to Pauli exclusive principle
for $s$-wave interaction, now becomes possible. The scatterings between the atoms near the Fermi points are essential for these pairings. To this end,
we diagonalize the single particle Hamiltonian as $\mathcal{H}(k)=\varepsilon_\pm(k) a_{\pm, k}^\dagger a_{\pm, k}$, where
\begin{eqnarray}
        \mqty(a_{\uparrow,k} \\ a_{\downarrow,k})=
            \mqty(a(k)&b(k)\\c(k)&d(k))
                \mqty(a_{+,k} \\ a_{-,k}),
                    \label{eq-ampeq}
\end{eqnarray}
and $a(k)$, $b(k)$, $c(k)$, $d(k)$ are coefficients that diagonalize $\mathcal{H}(k)$. In the weak interaction limit, we only need to consider the scatterings near
the Fermi surface, in which we may linearize the spectra using the essential idea of the Tomonaga-Luttinger liquid theory \cite{tomonaga_remarks_1950,luttinger_exactly_1963},
$\mathcal{H}(k) \simeq \sum_{k\sim k_F^1} \nu_F^1 (k-k_F^1) a_{+,k}^\dagger a_{+,k} +\sum_{k\sim k_F^2} \nu_F^2 (k-k_F^2) a_{+,k}^\dagger a_{+,k}$, with
$\nu_F^i=\pdv k\varepsilon_+(k)|_{k=k_F^i}$ denoting the Fermi velocity. In this way, the interaction can be rewritten as
\begin{eqnarray}
    V_{\text{int}}=-U\sum_{k_1,k_2,q}\mathcal{V}_{k_1,k_2}^q a^\dagger_{+,(k_1+q)} a^\dagger_{+,(k_2-q)} a_{+,k_2} a_{+,k_1},
            \label{eq-interactionexpand}
\end{eqnarray}
where $\mathcal{V}_{k_1,k_2}^q= a^*(k_1+q)c^*(k_2-q)c(k_2)a(k_1)$. We have four different scatterings, which, following the g-ology terminologies \cite{giamarchi2004quantum}, are 
written as: (1) Forward scattering on the same branch, $k_1\sim k_{F}^i$, $k_2\sim k_F^i$ (for $i$ = 1, 2), $q\sim 0$, $g_4^{L/R}= Ua^*(k_F^i)c^*(k_F^i)c(k_F^i)a(k_F^i)$;
(2) Dispersion scattering, $k_1\sim k_{F}^1$, $k_2\sim k_F^2$, $q\sim 0$, $g_2= Ua^*(k_F^1)c^*(k_F^2)c(k_F^2)a(k_F^1)$; and
(3) Backward scattering, $k_1\sim k_{F}^1$, $k_2\sim k_F^2$, $q\sim k_F^2-k_F^1$, $g_1= Ua^*(k_F^2)c^*(k_F^1)c(k_F^2)a(k_F^1)$.

\begin{figure}[htpb]
    \centering
    \includegraphics[width=0.49\textwidth]{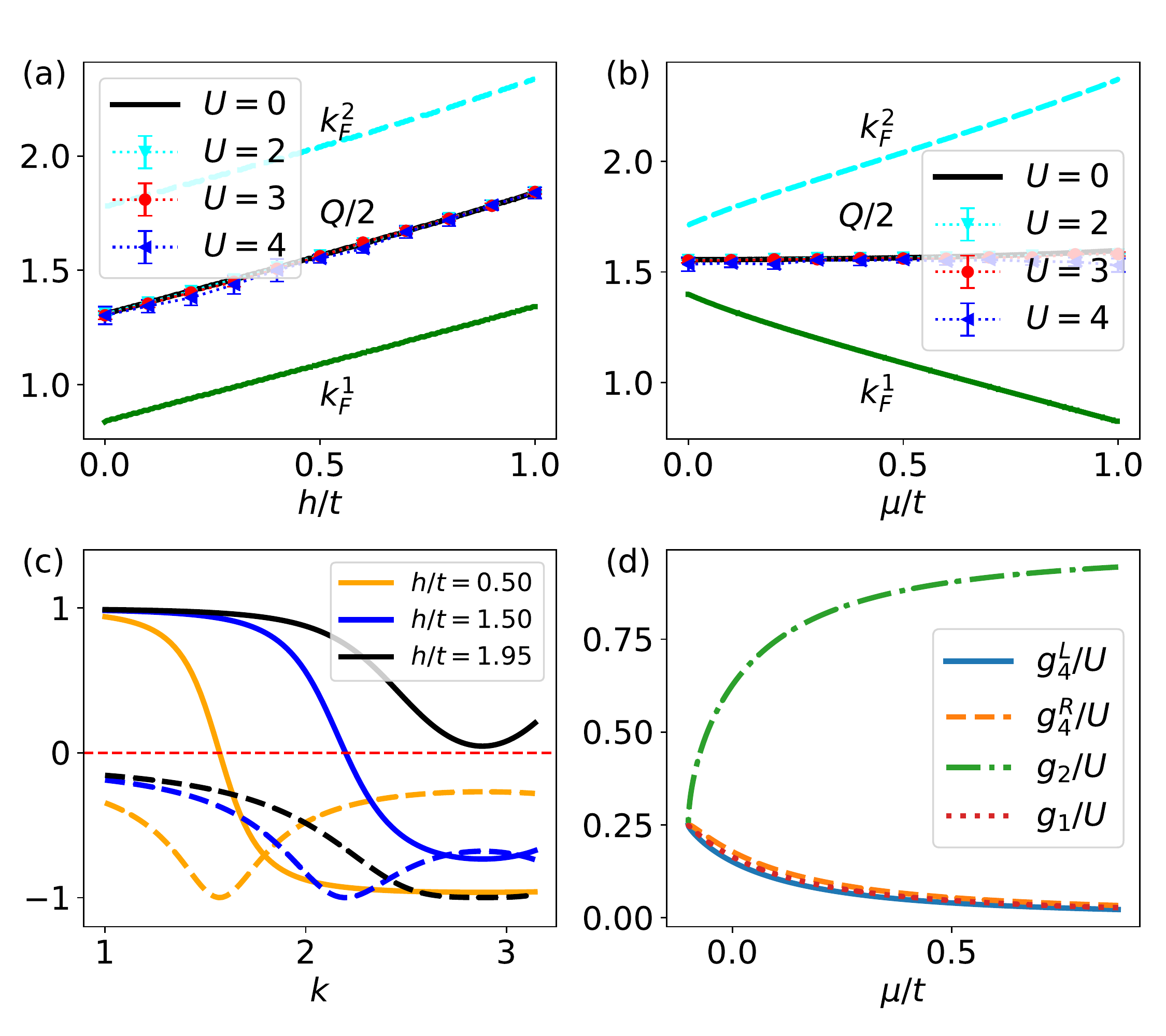}
    \caption{(Color online). The momentum $Q$ carried by the Cooper pairs and the two Fermi points $k_\text{F}^{1,2}$ 
    as a function of $h$ (a) and $\mu$ (b). The symbols are $Q/2$ calculated by mean field theory with finite $U$. 
    (c) Spin polarization $\langle \sigma_x\rangle $ (dashed lines) and $\langle \sigma_z\rangle$ (solid lines) as a function of $k$ for different 
    Zeeman field. (d) All possible scatterings in terms of g-ology, in which the dispersion scattering is dominated. Other parameters are $h/t=1.0$, 
    $\Omega/t=0.4$.}
    \label{fig-fig2}
\end{figure}

In these scatterings, only the dispersion
scattering $g_2$ between the right and left movers is important for pairing. The backward scattering $g_1$ is important only near half filling. In Fig. \ref{fig-fig2}a-b, 
we plot the evolution of $k_\text{F}^{1,2}$ and $Q/2$ as a function of $h$ and $\mu$, which exhibit good linearity over a wide range. However their spins change in a non-monotonously 
way with the increasing of momentum $k$ (see Fig. \ref{fig-fig2}c), which is totally different from that in the Rashba SOC model. 
We plot the evolution of $g$ parameters in 
Fig. \ref{fig-fig2}d. We find that over a wide range of chemical potential, the scattering is always dominated by the pairing term described by $g_2$ due to nearly
opposite spin polarization at the Fermi points. With the increasing of chemical 
potential, the other three parameters, $g_4^L$, $g_4^R$ and $g_1$, all decrease to very small amplitude and $g_2\sim U$. 
For comparison we have employed the same analysis to the model with Rashba SOC and find that for a sufficient large chemical potential and strong SOC strength, 
the maximum value of $g_2 \sim U/4$. This result indicates the observation of topological superfluids in our model with much weaker attractive interaction.

\begin{figure}[htpb]
    \centering
    \includegraphics[width=0.49\textwidth]{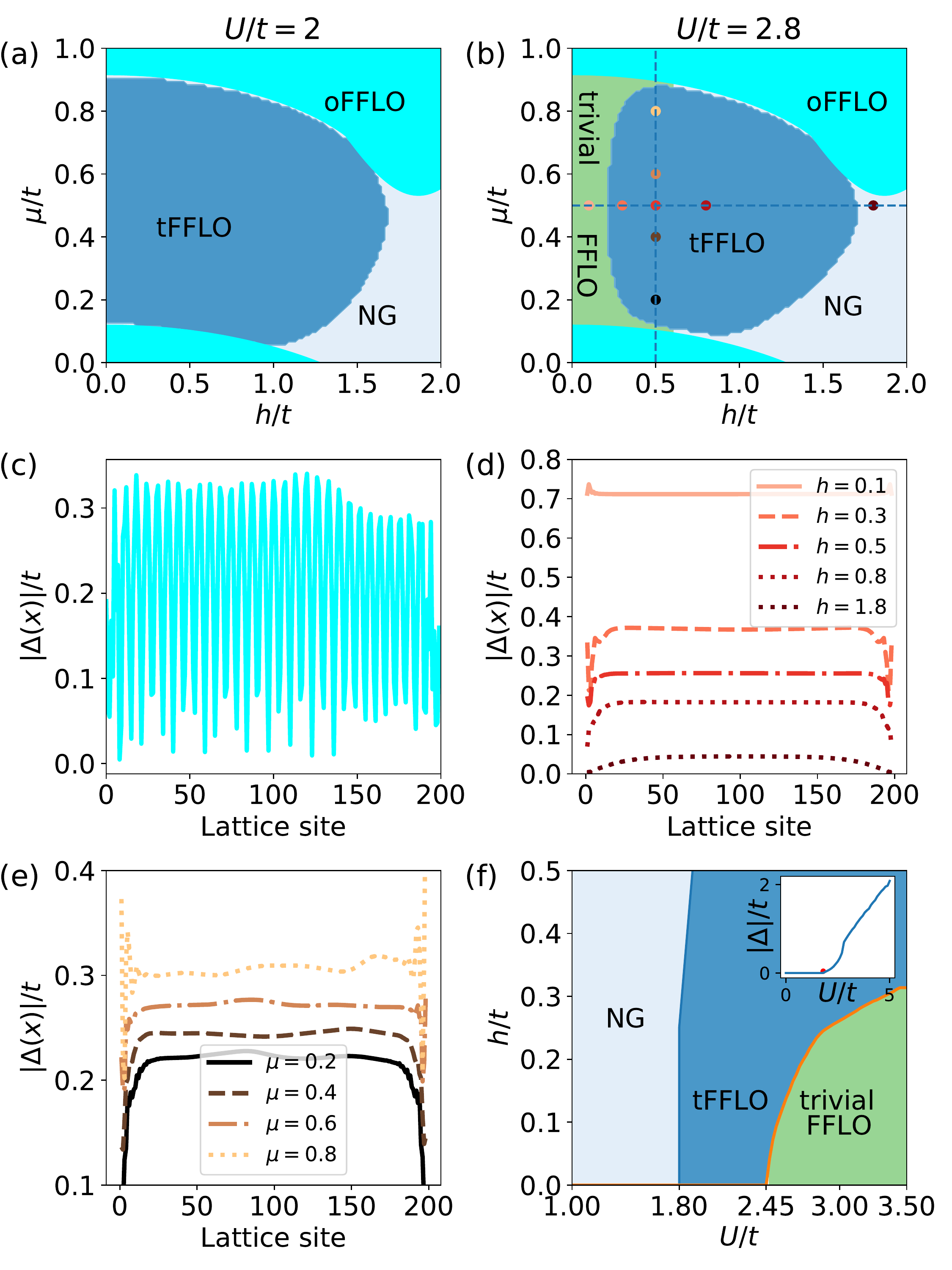}
    \caption{(Color online) Phase diagrams in the $\mu$ and $h$ plane for $U = 2.0t$ (a)
    and $U = 2.8t$ (b). (b) Oscillating order parameters in the oFFLO phase. (d) and (e) Magnitude of order parameters in the FFLO phase regime in a finite chain. These 
    two results correspond to the horizontal and vertical lines in (b). (f) Phase diagram in the $h$ and $U$ plane for $\mu = 0.5t$. Inset shows the order parameter as a 
    function of $U$, which indicates a finite threshold $U > U_c = 1.8t$ for pairing. }
    \label{fig-fig3}
\end{figure}

In following we underpin the above conclusions via the self-consistent mean-field theory in real space
\cite{qu_fulde-ferrell-larkin-ovchinnikov_2014}
\begin{eqnarray}
-Un_{m,\uparrow}n_{m,\downarrow} \simeq  \Delta_m a_{m,\uparrow}^\dagger a_{m,\downarrow} +\text{h.c.} +{\abs{\Delta_m}^2 \over U}, 
    \label{eq-meanfield}
\end{eqnarray}
where $\Delta_m = U \langle c_{m\uparrow} c_{m\downarrow}\rangle$, the pairing strength, are solved self-consistently.
We do not consider the Hartree term, which can be included in the chemical potential. This model may support a uniform FFLO phase 
in some proper parameter regime with (see Fig. \ref{fig-fig1}c), 
\begin{equation}
    \Delta(x) = \Delta_0 e^{iQ x}.
\end{equation}
In Fig. \ref{fig-fig2}a-b, the symbols are the calculated $Q$ from mean-field theory for different $U$, which always yield 
$Q = k_\text{F}^1 + k_\text{F}^2$. A rough estimation shows that for $U \sim 2.0t$, the magnitude of
pairing strength is $|\Delta| \sim 0.2 - 0.5t$. This value is one order of magnitude larger than that generated by Rashba SOC \cite{qu_fulde-ferrell-larkin-ovchinnikov_2014}, 
in which a comparable pairing need a much bigger interaction in the intermediate or strong coupling regime ($U \sim 4.5 - 5.0t$).
However, it was shown in \cite{liang2015unconventional} that with this intermediate or strong coupling,
unconventional pairings may become important in some parameter regimes. 
This result is consistent with the dominated dispersion scattering in Eq. \ref{eq-interactionexpand}.

In this case the corresponding Bogoliubov-de Gennes (BdG) equation can be written as
\begin{eqnarray}
    H_{\text{BdG}}=\mqty( H_0(k) & i\sigma_y \Delta_0 e^{iQx} \\
    -i\sigma_y \Delta_0^\dagger e^{-iQx} & -H_0^*(k)),
    \label{eq-HBdG}
\end{eqnarray}
where $k = -i\partial_x$. The global phase carried by the order parameter can be gauged out by an unitary
transformation, $\mathcal{U} = \text{diag} (e^{iQx/2}, e^{-iQx/2})$, while yields
\cite{chan_pairing_2014,qu_topological_2013},
\begin{eqnarray}
    \mathcal{U}^\dagger H_{\text{BdG}} \mathcal{U}=
    \mqty( H_0(k+Q/2) & i\sigma_y \Delta_0 \\ -i\sigma_y \Delta_0^\dagger & -H_0^*(k-Q/2)).
    \label{eq-mdagerhm}
\end{eqnarray}
Since the Zeeman field in the off-diagonal term is independent of $k$, the momentum shift $Q/2$ will not enter the off-diagonal term.
This is totally different from the model with Rashba SOC, in which $Q$ will contribute to an effective in-plane Zeeman field and induce 
a strong tilting effect to the band structures\cite{chan_pairing_2014}. The topological phase boundary for
the above model takes place at $k = 0$ and $\pi$ are given by \cite{kitaev_unpaired_2001},
\begin{eqnarray}
    \text{pf}(H(k)\sigma_x) && = \Omega^2 - \Delta_0^2- \left[ h-\mu-2t\cos(k+\phi) \right] \nonumber\\
    &&\times \left[ -h-\mu-2t\cos(k) \right], \quad k=0, \quad \pi.
    \label{eq-pf}
\end{eqnarray}
This index $\nu = \text{sgn}(\text{pf}(H(k))\sigma_x)$ is used to characterize the topological invariant, in which 
$\nu = +1$ (-1) corresponds to the trivial and topological phase, respectively.

We plot the phase diagram as a function of $U$ and $h$ in Fig. \ref{fig-fig3}, focusing on the inverted band case with $\phi = 7\pi/6$, 
stimulated by the recent experiments in Refs. \cite{wall_synthetic_2016,kolkowitz_spinorbit-coupled_2017}. The similar phases
can be found for other magnetic flux. We find a large parameter regime for topological FFLO phase (denoted as tFFLO), in which the magnitude of pairing, $|\Delta_m|$, is 
almost uniform as a function of $h$ and $\mu$ (see Fig. \ref{fig-fig3}d-e). When $U = 2.0t$, we find that all the FFLO states are topological 
nontrivial, which means that even without out-of-plane field $h$, the system can still be a topological phase. This observation is
consistent with Eq. \ref{eq-interactionexpand}, in which not only the gauge potential but also the momentum $Q$ carried by the Cooper pairs contribute to
the topological boundaries. For $U = 2.8t$, we may have a trivial FFLO phase, which is also fully gapped. This fully gapped phase is attributed
to $Q = k_\text{F}^1+k_\text{F}^2$ with weak interaction, which ensures that after a parallel move of momentum by $\pm Q/2$ 
for the particle and hole Hamiltonian, their Fermi points are exactly coincide in the Fermi surface, at which point an sizable gap 
is opened by the order parameter. We can even prove exactly that the system is always fully gapped when $Q$ is determined by the two Fermi points.

\begin{figure}[htpb]
    \centering
    \includegraphics[width=0.49\textwidth]{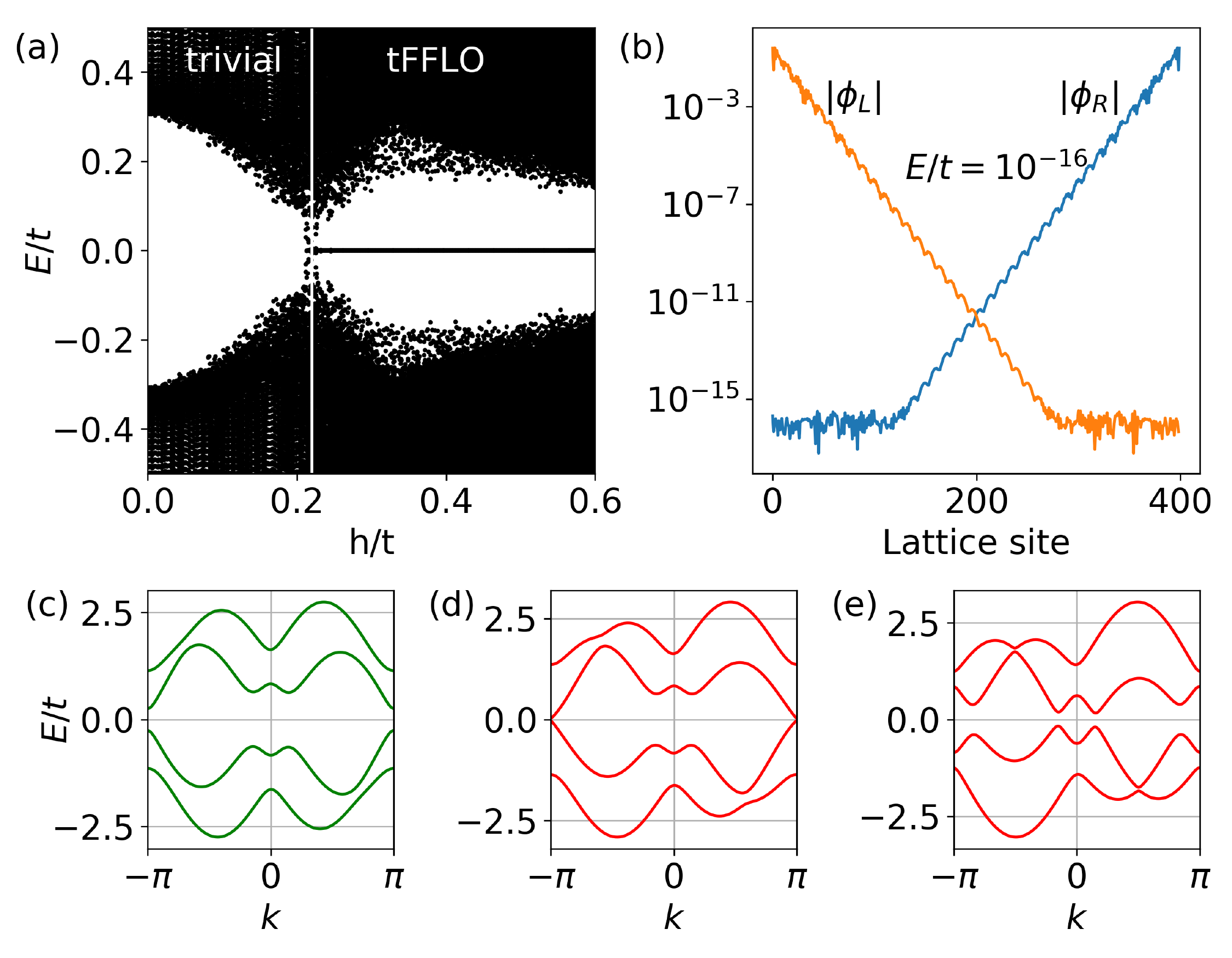}
    \caption{(Color online) Majorana zero modes in an one dimensional chain induced by artificial gauge potential. The phase boundary between 
    trivial FFLO and tFFLO is determined by the appearance of edge modes and $\nu = -1$ (Eq. \ref{eq-pf}). Parameters are $\mu = 0.5t$, $U = 2.8t$.
    (b) Wave functions of the edge modes for $h = 0.4t$, $\mu = 0.2t$. (c) - (e) Band structures of the BdG equation for $\mu = 0.5t$ and $h = 0.1t$ (trivial, $\nu=+1$), $0.22t$ (critical) and $0.5t$ (tFFLO, $\nu=-1$), respectively.}
    \label{fig-fig4}
\end{figure}

It is also possible to observe oscillating FFLO phase (denoted as oFFLO), in which the order parameters oscillate fast in real space due to the competition of 
two different momenta carried by the Cooper pairs, which can be realized in the small or large chemical potential limit with four Fermi points (
see also Fig. \ref{fig-fig1}b). For weak interaction $U$, the boundaries between tFFLO and oFFLO can be determined by the single particle band structures in 
Fig. \ref{fig-fig1}b-c. The phase diagram as a function of $h$ and $U$ is presented in Fig. \ref{fig-fig3}f, in which the inset shows a finite 
threshold $U > U_\text{c} \simeq 1.8t$ is required to realize pairing. We have numerically verified that this threshold is roughly one half of that in the 
model with Rashba SOC \cite{qu_fulde-ferrell-larkin-ovchinnikov_2014,qu_topological_2013,potirniche_superconductivity_2014}.
For large enough Zeeman field the pairing is completely destroyed in the normal gas (denoted as NG) phase (Fig. \ref{fig-fig3}a-b).

We finally discuss the appearance of Majorana zero energy modes at the two ends in the tFFLO phase regime. In the trivial FFLO phase 
regime, the system is still fully gapped (see the band structure in Fig. \ref{fig-fig4}c). With the increasing of $h$, the gap is 
closed and reopened at $k = \pi$ (Fig. \ref{fig-fig4}d), and then the edge modes appear in the tFFLO phase. The wave function of the Majorana zero modes 
are shown in Fig. \ref{fig-fig4}b. 

To conclude, in this work we propose to realize topological superfluids and associated Majorana zero modes based on artificial gauge potential, 
which can be regarded as a ``complementary'' mechanism to the Rashba SOC. This new method has the advantage of realizing 
the topological phases with much weaker attractive interaction due to the dominated dispersion scattering near the Fermi surface. We 
find that all the topological FFLO superfluids are fully gapped, and we map out the corresponding phase diagram using self-consistent mean-field
theory. Our model has the potential to be the first experimental system to realize the long-sought FFLO phase and Majorana zero modes, in regarding of 
the above mentioned advantages and the negligible role of heating effect in alkaline and rare-earth atoms for the artificial gauge potentials.

\textit{Acknowledgements.} M.G. is supported by the National Youth Thousand Talents Program (No. KJ2030000001), the USTC start-up funding (No. KY2030000053), 
the NSFC (No. GG2470000101). Z. Z. is supported by the Young Scientists Fund of the National Natural Science Foundation of China (Grant No. 11704367). M.G. thank 
Prof. W. Yi for valuable discussion.

\bibliography{ref}
\end{document}